\documentclass[12pt,a4paper,dvips]{article}
\usepackage{a4p}
\usepackage{times,psmath}
\usepackage{cite,mcite}
\usepackage{graphicx}
\usepackage{physics }
\usepackage{ams_title,ifthen,Lep}
\journalname{Phys. Lett. B}
\date{}
\myjournaln
\graphicspath{{.}}
\newlength{\capindent}
\newlength{\capwidth}
\newlength{\figwidth}
\newcommand{\icaption}[2][!*!,!]{\hspace*{\capindent}%
  \begin{minipage}{\capwidth}
    \ifthenelse{\equal{#1}{!*!,!}}%
      {\caption{#2}}%
      {\caption[#1]{#2}}
  \end{minipage}}
\newcommand{\He}{\ensuremath{\mathrm{He}}}
\newcommand{\aHe}{\ensuremath{{\,\overline{\mathrm{\!He\!}}\,}}}
\newcommand{\R}{\ensuremath{{\mathrm{R}}}}
\begin{document}
\begin{titlepage}
\title{Search for Antihelium in Cosmic Rays}
\author{The AMS Collaboration}
%
% The abstract
%
\begin{abstract}
The Alpha Magnetic Spectrometer (AMS) was flown on the space shuttle
{\it Discovery} during flight STS--91
in a 51.7$^\circ$ orbit at altitudes between 320 and 390\,km.
A total of $2.86\times 10^6$ helium nuclei 
were observed in the rigidity range 1 to 140\,GV.
No antihelium nuclei were detected at any rigidity.
An upper limit on the flux ratio of antihelium to helium
of $< 1.1\times 10^{-6}$ is obtained.
\end{abstract}
%
% Adds "To be submitted to ..." or "Submitted to ...", if relevant
%
\submitted
\end{titlepage}
%
%%%%%%%%%%%%%%%%%%%%%%%%%%%%%%%%%%%%%%%%%%%%%%%%%%%%%%%%%%%%%%%%%%%%%%%%%%%%%%%
% Introduction
%%%%%%%%%%%%%%%%%%%%%%%%%%%%%%%%%%%%%%%%%%%%%%%%%%%%%%%%%%%%%%%%%%%%%%%%%%%%%%%
%

\section*{Introduction}
The existence (or absence) of antimatter 
nuclei in space
is closely connected with the
foundation of the theories of 
elementary particle physics,
CP--violation, 
baryon number nonconservation, 
Grand Unified Theory (GUT), 
etc.
Balloon--based 
cosmic ray searches 
for antinuclei
at altitudes up to 40\,km
have been carried out for more than 20
years; all such searches have been negative~\mcite{smoot75,
steigman76,badhwar78,buffington81,golden97,ormes97,saeki98}.
The absence of annihilation gamma ray peaks excludes the presence
of large quantities of antimatter within a distance of the order of 10\,Mpc
from the earth.
The baryogenesis models are not yet supported by particle physics
experimental data.
To date baryon nonconservation and large levels of CP--violation
have not been observed.
The Alpha Magnetic Spectrometer (AMS)~\cite{nimproposal} is scheduled
for a high energy physics 
program on the International
Space Station.  
In addition to searching for dark matter
and the origin of cosmic rays,
a major objective of this program 
is to search for antinuclei
using an accurate, large acceptance
magnetic spectrometer.
AMS was flown on the
space shuttle {\it Discovery} on flight STS--91 in June 1998.
This was primarily a test flight that would enable the AMS team
to gather data on background sources,
adjust operating parameters and
verify the detector's performance under actual space flight conditions.
A search for antihelium nuclei 
using the data collected during this precursor flight is reported.
The signal investigated is nuclei with charge $Z=-2$.

\section*{AMS on STS--91}
\begin{figure}[ht]
  \begin{center}                     % my sketch.
    \includegraphics[width=\figwidth]{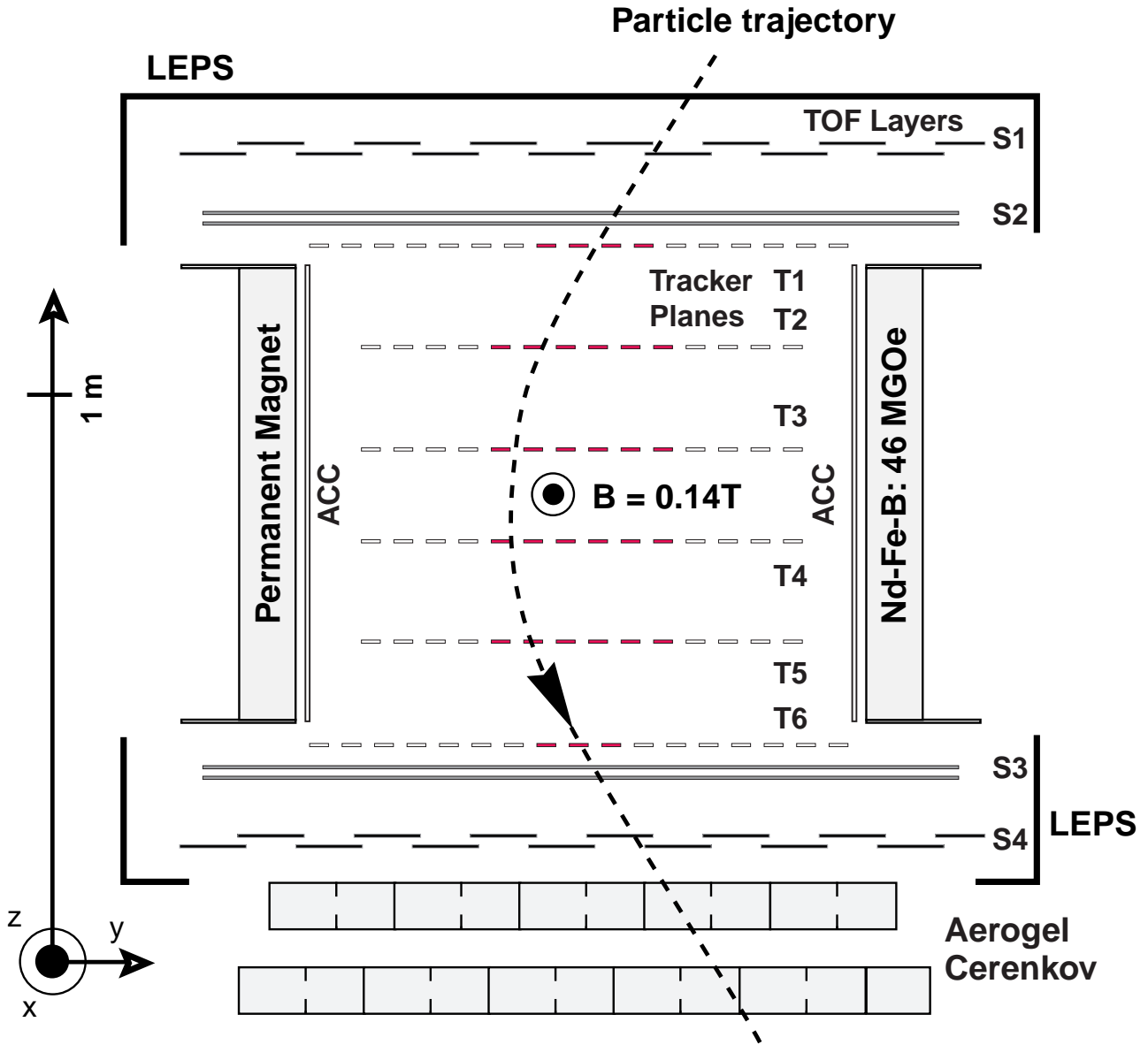}
  \end{center}
  \caption{Schematic view of AMS as flown on STS--91 showing
the cylindrical permanent magnet, 
the silicon microstrip tracker planes T1 to T6, 
the time of flight (TOF) hodoscope layers S1 to S4,
the aerogel cerenkov counter,
the anticoincidence counters (ACC) and 
low energy particle shields (LEPS).}
  \label{schema}
\end{figure}

A schematic cross section 
in the bending plane of AMS as flown on STS--91, Fig.~\ref{schema},
shows the permanent magnet, tracker,
time of flight hodoscopes, Cerenkov counter and anticoincidence counters.
The AMS coordinate system, as shown, coincides with the shuttle
coordinate system, with the z--axis 
(up in the figure) pointing
out of the shuttle payload bay and the x--axis
pointing towards the tail of the shuttle.
The geometric acceptance was $\sim 0.3$\,m$^2$sr.
AMS as flown on STS--91 will be described in detail 
elsewhere~\mcite{ams_00,*viertel98}.

The magnet provided the analyzing power of the spectrometer. 
It was made of 1.9 tons of Nd-Fe-B
in the shape of a cylindrical shell 
of inner diameter 1115\,mm and length 800\,mm.
The  Nd-Fe-B was magnetized to 46\,MGOe 
with the direction varying 
to provide a dipole field in the x direction,
perpendicular to the cylinder axis.
At the center the magnetic field was 0.14\,Tesla and 
the analyzing power, $BL^2$, was 0.14\,Tm$^2$.

\begin{figure}[tb]
  \begin{center}                                  
    \includegraphics[width=\figwidth]{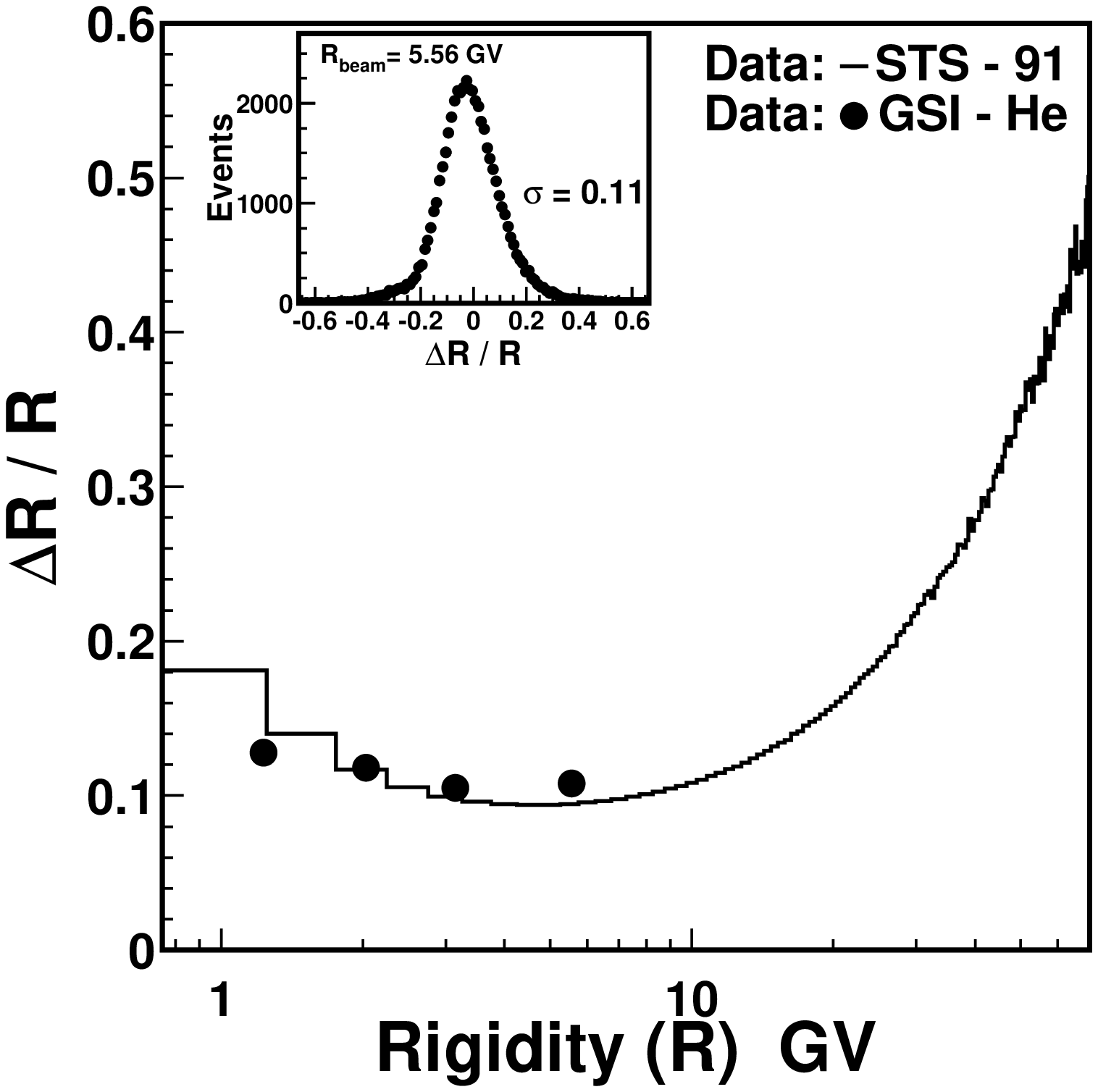}
  \end{center}
  \caption{Rigidity resolution 
            for $|Z|=2$ flight data (histogram) 
compared with the GSI He test beam (points).\newline
Inset: Typical rigidity resolution,
$\Delta \R/\R = (\R_{measured}-\R_{beam})/\R_{beam}$,
from the GSI He data. 
}
  \label{aheresol_before}
\end{figure}

The trajectory of charged particles traversing the magnet bore
was observed with a tracker made of six planes, T1 to T6, of double sided 
silicon microstrip detectors~\mcite{tracker}.
For AMS on STS--91 half of the tracker area was equipped. 
From the deflection the rigidity, $\R=pc/|Z|e$\,(GV), was measured.
The tracker also provided a determination of charge magnitude, $|Z|$,
through multiple energy loss measurements.
Special care was taken to minimize the amount of material  
in the tracker construction;
the total amount of material 
within the tracker volume
was less than
3\% of a radiation length parallel to the z--axis.
The tracker alignment was made first with metrology,
continuously monitored with an infrared laser system 
 and then verified with high
momentum tracks from the CERN PS test beam.
During flight
hits in the tracker were measured with
an accuracy of $\sim 10\,\mu$m in the bending, or 
y, direction 
and $\sim 30\,\mu$m in the x and z directions.
The resolution in terms of rigidity
was verified for $|Z|\ge2$ nuclei
using helium and carbon ion beams at GSI--Darmstadt.
Fig.~\ref{aheresol_before} 
shows the rigidity resolution 
for
$Z=2$ flight data and the agreement
with the $Z=2$ helium data measured at GSI.
Note that 
at low momenta the resolution was limited by multiple scattering.

\begin{figure}[ht]
  \begin{center}                                % fixed up.
    \includegraphics[width=\figwidth]{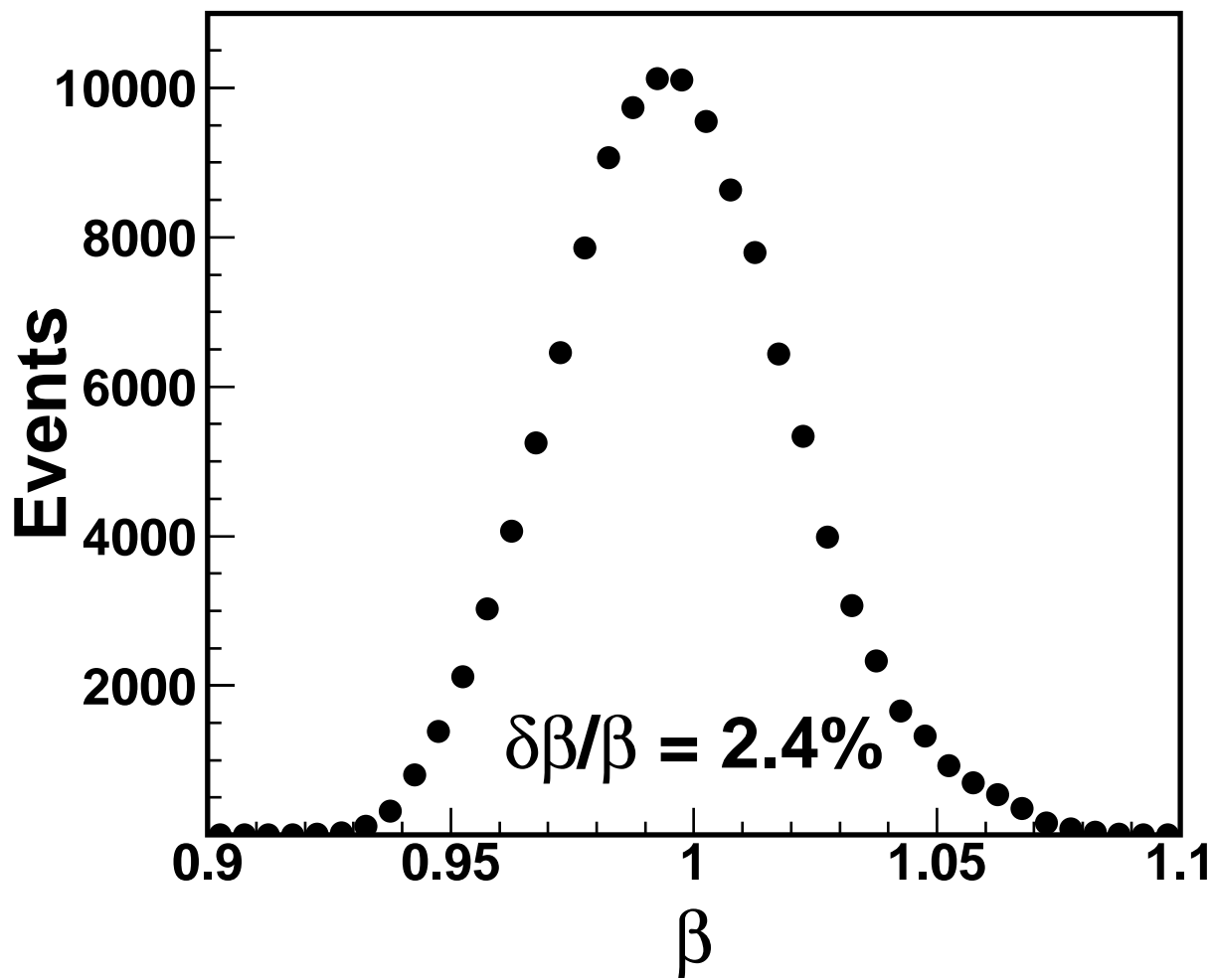}
  \end{center}
  \caption{Measured velocity, $\beta=\mathrm{v}/c$, distribution for 
          $|Z|=2$ events with $\R>16\,\mathrm{GV}$.}
  \label{betares}
\end{figure}

The particle direction and velocity were measured with a four layer,
S1 to S4,
time-of-flight (TOF) hodoscope. 
Each layer consisted of 14 scintillator paddles of 
thickness 10\,mm, width 110\,mm,
hermetically arranged with a 5\,mm overlap. 
As shown in Fig.~\ref{schema}, two layers were
above the magnet and two below.
The paddles in each pair were orthogonal.
The pulse height information recorded from the TOF paddles provided an 
additional determination of $|Z|$. 
The typical accuracy of the time of flight measurements was 105\,psec
for $|Z|=2$. 
Fig.~\ref{betares} shows the velocity,
$\beta = \mathrm{v}/c$, resolution for 
high rigidity $|Z|=2$ particles.

The velocity measurement was complemented by a threshold Cerenkov
counter made of aerogel with a refractive index of 1.035. 

A layer of anticoincidence scintillation counters (ACC) covered the inner
surface of the magnet to reject the background caused by
particles passing through or
interacting in the  magnet walls and support structures. 
The detector was also shielded
from low  energy (up to several MeV) particles
by thin carbon fiber walls (LEPS).
For particles arriving from above, as shown in Fig.~\ref{schema},
the amount of material at normal incidence 
was $1.5\,\mathrm{g/cm}^2$ in front of the TOF system,
and $3.5\,\mathrm{g/cm}^2$ in front of the tracker.

During construction, the detector components went though
extensive space qualification tests
(acceleration, vibration, thermal vacuum, electromagnetic interference
and radiation).
For example, the magnet was tested in a centrifuge to 17.7\,g.
Key electronics components were tested at Dubna in heavy ion beams of 
Ne, Ar and Kr.

During flight the detector was located in the payload bay of 
the space shuttle and operated in vacuum.
Events were triggered by the coincidence of signals in all four
TOF planes consistent with the passage of a charged particle
through the active tracker volume.  
Triggers with a coincident signal from the ACC were vetoed.
The detector performance as well as 
temperature and magnetic field 
were monitored continuously.  
A total of 100 million triggers were recorded.

After the flight, the detector was checked again:
\begin{itemize}
\item first, the detector was placed in a heavy ion (He, C) beam 
      from 1.0 to 5.6\,GV at 600 different incident angles.
      This test was done with a total of 45 million events and
      was carried out at GSI--Darmstadt.
\item second, the detector was placed in a proton and pion beam
      at CERN with momentum from 2 to 14\,\GeV\ at 1200 different
      incident angles.  This test was done with a total
      of 100 million events.
\end{itemize}
The continued monitoring of the detector confirmed that the
detector performance 
before, during and after
the flight remained the same.
In particular, the alignment of the silicon tracker remained the same
to an accuracy of $\sim 5\,\mu$m.

\section*{Event Selection}
After the shuttle had attained orbit,
data collection commenced  
on  3 June 1998      
and continued over the next nine days
for a total of 
184 hours.
During data taking the shuttle altitude varied from 
320 to 390\,km
and the latitude ranged between 
$\pm$ 51.7\,degrees.
Before rendezvous with the MIR space station  
the attitude of the shuttle  
was maintained to keep the z--axis of AMS 
(see Fig.~\ref{schema})
pointed within 45 degrees of the zenith.
While docked, the attitude was constrained by MIR requirements
and varied substantially.
After undocking the pointing was maintained within
1, 20 and then 40 degrees of the zenith.
Shortly before descent the shuttle turned over
and the pointing was towards the nadir.
For this search, 
data collected while passing through the South Atlantic
Anomaly was excluded.

\newpage
The procedure to search for antihelium began
with event reconstruction, which included:
\begin{itemize}
\item Measurement of the particle rigidity, $\R$, from
      the deflection of the trajectory measured by the tracker
      in the magnetic field.  To ensure that the particle
      was well measured, 
      hits in at least four tracker planes were required
      and the fitting was performed with two different
      algorithms, the results of which were required to agree.
\item Measurement of the particle velocity, $\beta$, 
      and direction, $\hat\mathrm{z} =\:\pm 1$, from the TOF,
      where $\hat\mathrm{z} =\:- 1$ signifies a downward
      going particle in Fig.~\ref{schema}.  
\item Determination of the magnitude of the particle charge, $|Z|$,
      from the measurements of energy losses
      in the TOF counters and tracker planes (corrected for $\beta$).
\end{itemize}
From this reconstruction the sign of the particle charge
was derived from the deflection in the rigidity fit
and the direction. 
The particle mass was derived from $|Z|\R$ and $\beta$.

The major backgrounds to the antihelium ($Z=-2$) search
are the abundant amount of protons and electrons ($|Z|= 1$)
and helium ($Z=+2$).
To distinguish antihelium from e$^-$, p and $\He$, 
the detector response to  e$^-$, p and $\He$ was studied in
three ways:
\begin{itemize}
\item[(i)] from the  e$^-$, p and $\He$ data collected in flight.
\item[(ii)] from the $\He$ beam data at GSI and the p beam data at the CERN PS.
\item[(iii)] from Monte Carlo studies of (i) and (ii). 
\end{itemize}

\begin{figure}[ht]
  \begin{center}                                 % fixed up
    \includegraphics[width=\figwidth]{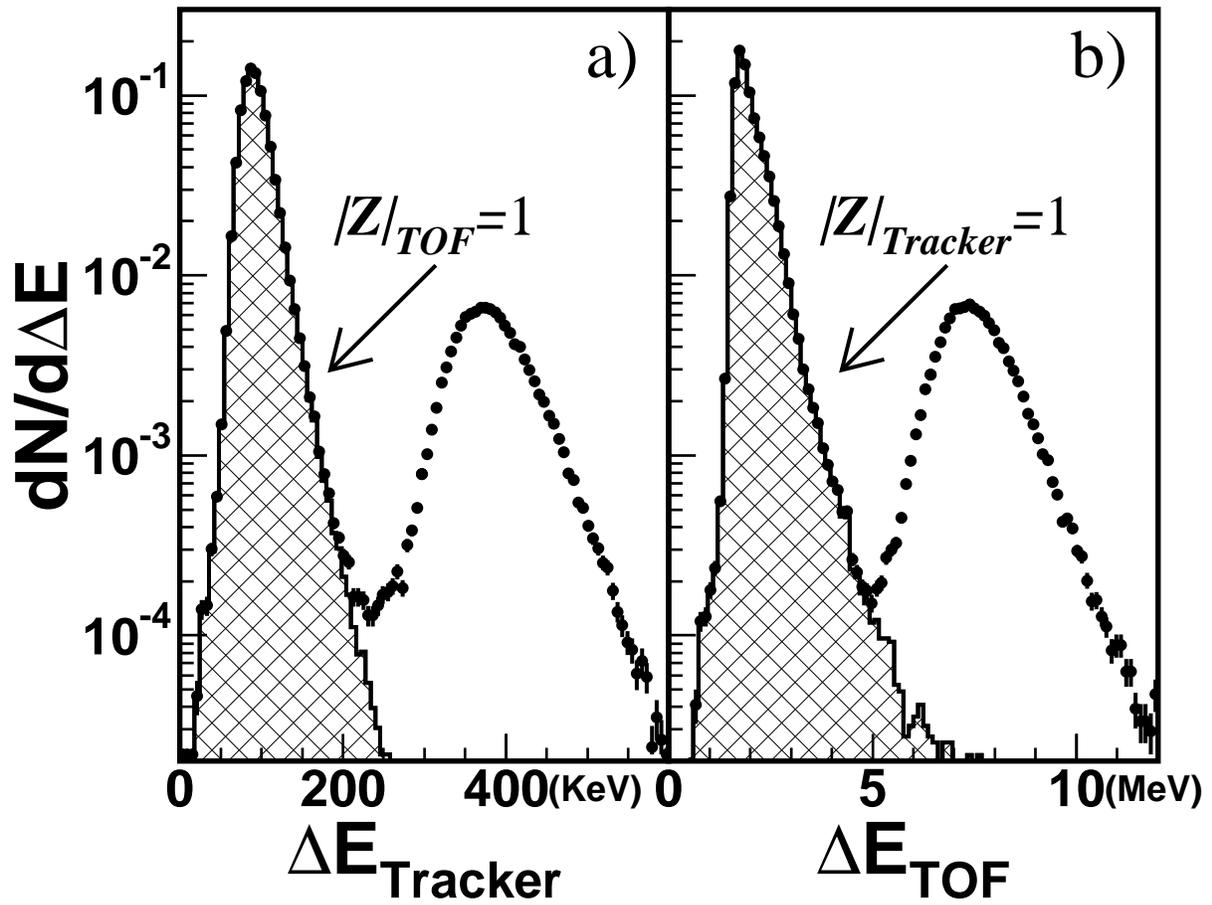}
  \end{center}
  \caption{Energy loss measurements (points) are made independently in
           the tracker (a) and TOF (b) for
       %    the first million 
           $|Z|\le2$ events.
           The hatched histogram
           shows which events were assigned 
           to be $|Z|=1$ by the other detector.}
  \label{chargeconf}
\end{figure}

\begin{figure}[ht]
  \begin{center}                                % fixed up.
    \includegraphics[width=\figwidth]{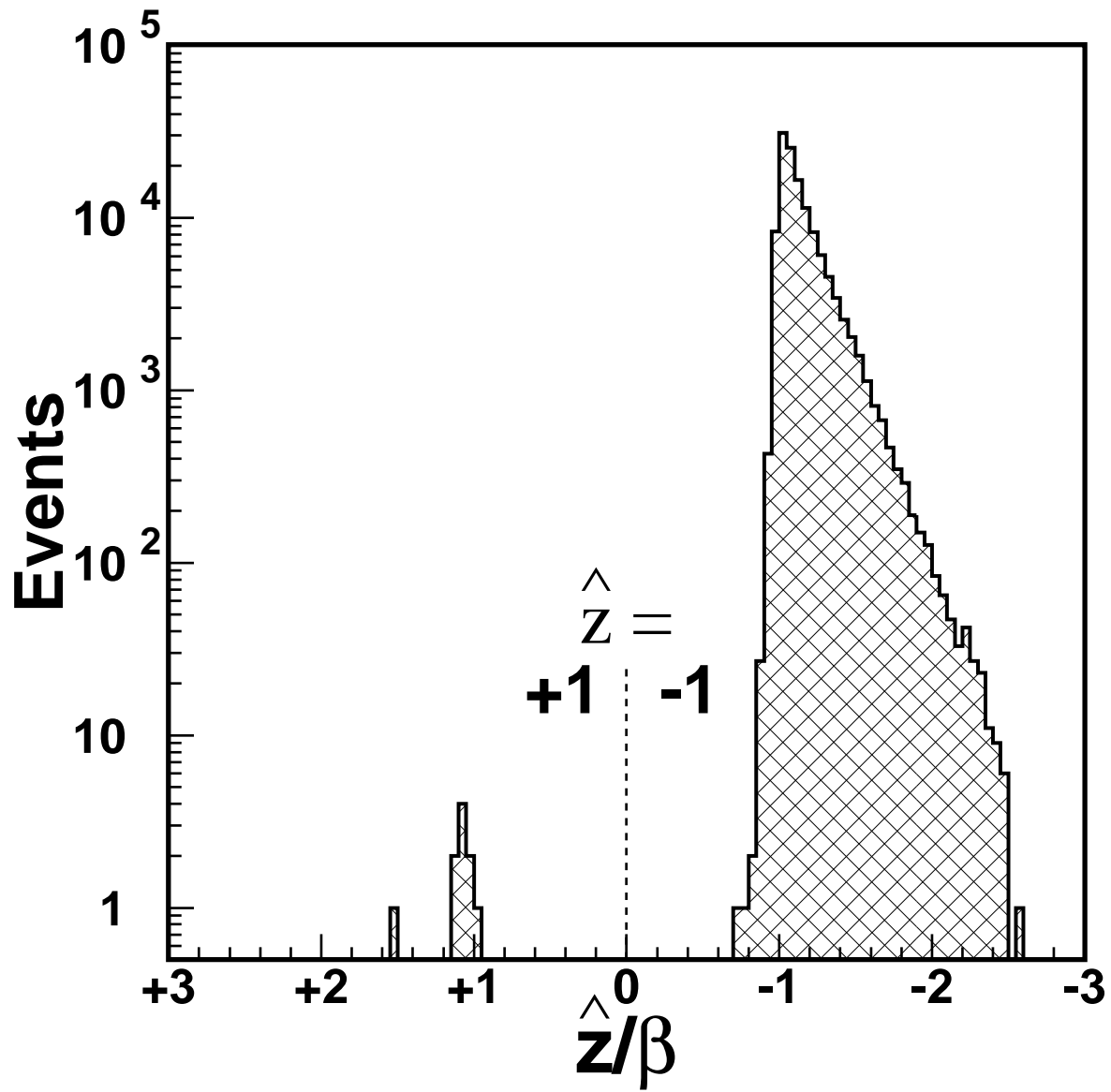}
  \end{center}
  \caption{A typical direction, $\hat{\mathrm z}/\beta$, distribution for 
           $|Z|=2$ events.
           As seen, the   
     $\hat\mathrm{z}=+1$ (or upward) 
    and  $\hat\mathrm{z}=-1$ (or downward) populations
    are clearly separated.}
  \label{twobeta}
\end{figure}

\begin{figure}[ht]
  \begin{center}                     % aka z2pic6
    \includegraphics[width=\figwidth]{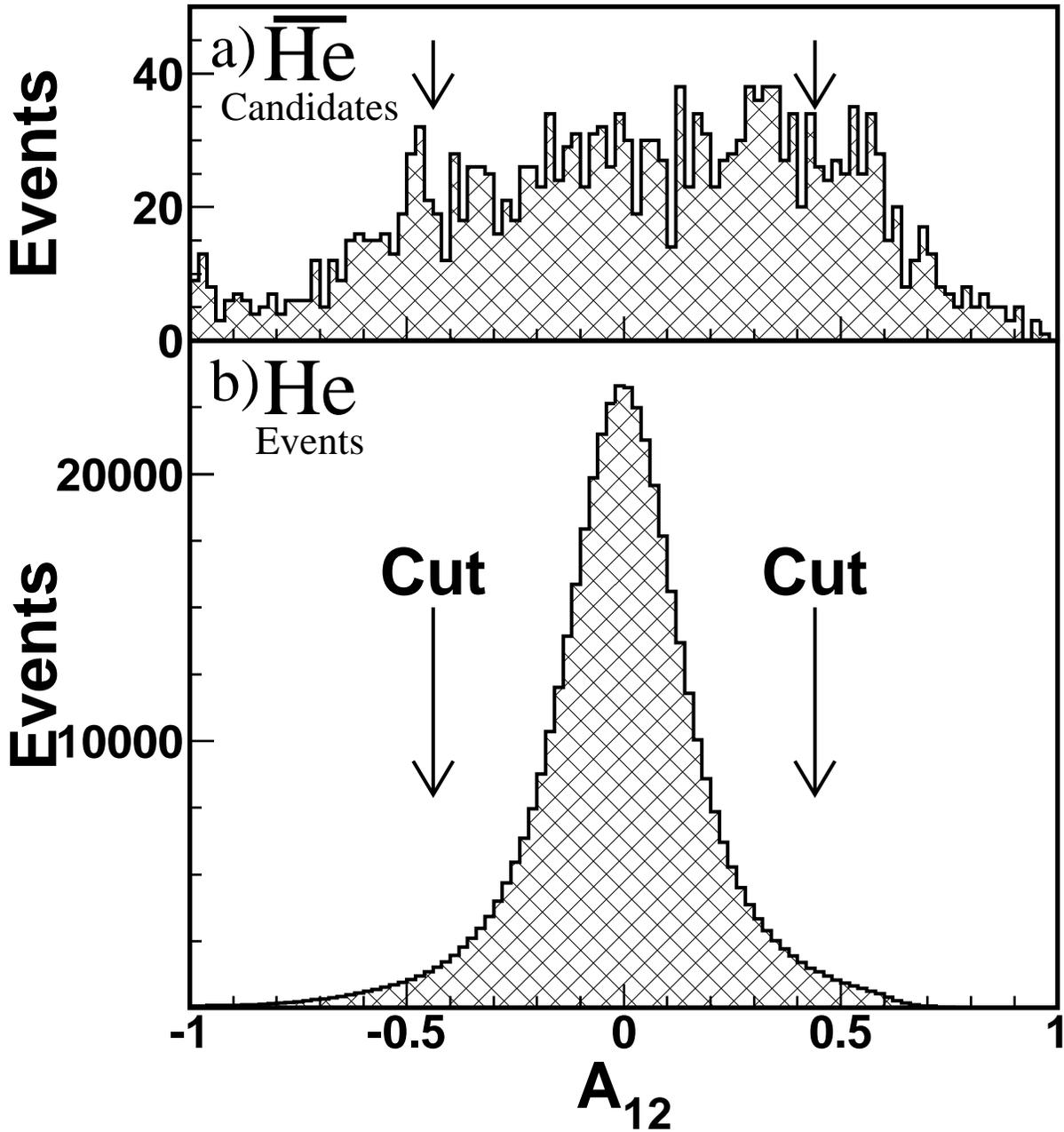}
  \end{center}
  \caption{The asymmetry, A$_{12}=(\R_1-\R_2)/(\R_1+\R_2)$, 
           of the rigidity measurements using the first, $\R_1$, 
           and last, $\R_2$, 
           three hits along the track for 
           $|Z|=2$ events.  Also shown are the cuts used.
           As seen the A$_{12}$ cuts reject much of the large angle scattering
           events (a).  The cuts do not reject the genuine signal (b).}
  \label{trassy}
\end{figure}

\begin{figure}[ht]
  \begin{center}                                
    \includegraphics[width=\figwidth]{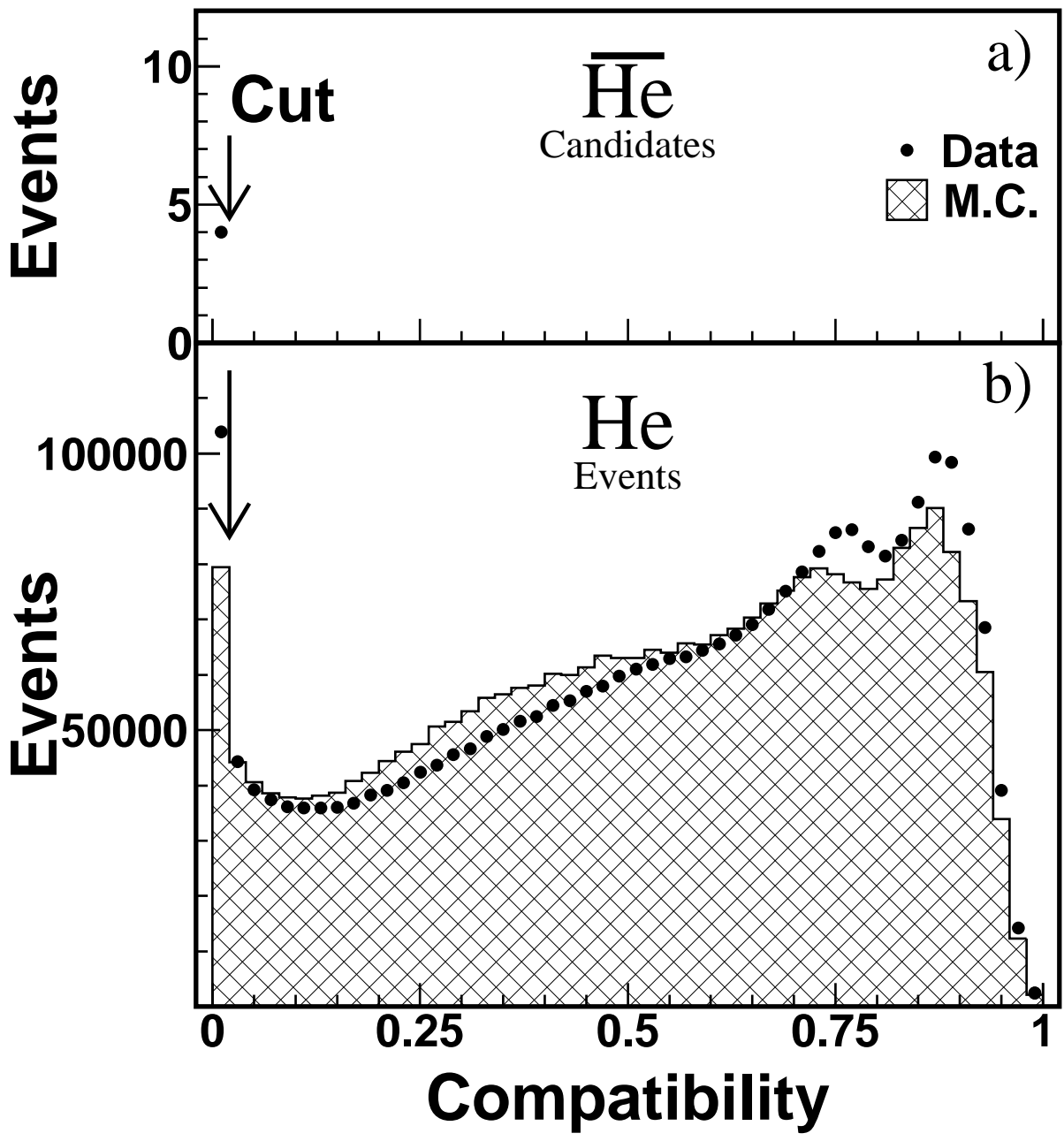}
  \end{center}
  \caption{(a) Compatibility of the measured event parameters,
           $\beta$, $\R$ and $|Z|$, to be an  
           $\aHe$ nucleus.\newline
           (b) Compatibility to be a $\He$ nucleus. 
           The hatched histogram 
           is the Monte Carlo prediction for $\He$ nuclei.}
  \label{probpart}
\end{figure}

Key points in the selection for $\aHe$ events and the rejection
of background were:

\ \newline
{\bf to select events with \boldmath $ |Z|=2$:} 
This was to ensure no contamination from
      $|Z|=1$ events with a wrongly measured charge magnitude which would
      mimic $|Z|>1$ events.
      Fig.~\ref{chargeconf} shows the energy deposition and the 
      assigned charge magnitude as measured
      independently by the TOF and the tracker.
      The probability of the wrong charge magnitude
      being assigned by the combined TOF and tracker measurements  
      was estimated to be less than $10^{-7}$.

\ \newline
{\bf to determine the sign of \boldmath $|Z|=2$ events:}
This was to distinguish $\He$ from $\aHe$.
This was done with the following method:
\begin{itemize}
\item[(i)] {\bf Identify the particle direction:}
      measurement of the particle direction 
      leads to 
      the correct assignment of the 
      sign of the charge.
      Fig.~\ref{twobeta} shows the particle direction,
      $\hat{\mathrm z}/\beta$, distribution.
      No events were observed between the $\hat{\mathrm z}= +1$ and 
      $\hat{\mathrm z}= -1$ populations which indicates there was 
      no leakage of particles from one population to the other
      and the direction was always correctly assigned.
\item[(ii)] {\bf Identify large angle nuclear scattering events:}
      events in which a single nuclear scattering 
      in one of the inner tracker planes, 
      T2--T5, introduced a large angle kink in the track and 
      might cause an incorrect measurement of the charge sign.
      This background was suppressed by a cut on the 
      estimated rigidity error.
      Additional suppression was achieved by requiring agreement for the 
      rigidity and charge sign 
      measured using all the hits in the tracker and 
      separately in the first three hits and 
      the last three hits along the track.
      Fig.~\ref{trassy} shows the asymmetry, 
      $ A_{12} = (\R_1-\R_2)/(\R_1+\R_2) $,
      of the rigidity measured with   
      the first and last three hits along the track,
      $\R_1$ and $\R_2$, and the cuts applied.
      From Fig.~\ref{trassy} we see that whereas these cuts 
      reject much of the large angle scattering events 
      (Fig.~\ref{trassy}a),
      the cuts do not reject the genuine signal (Fig.~\ref{trassy}b).
\item [(iii)]{\bf Identify events with collinear delta rays:}
      events with collinear debris, \eg\ delta rays, 
      from an interaction of the 
      primary particle in the tracker material which may shift a measured 
      point from the trajectory, leading to
      an incorrectly measured rigidity and charge sign.
      This background was efficiently rejected by
      an isolation cut which rejected events with an excess of
      energy observed  within 5\,mm of the track. 
\end{itemize}
A probabilistic function was then constructed
from  measurements of the velocity, rigidity and energy loss
which described the compatibility of these measurements
with the passage of a helium or antihelium nucleus of mass
$A=3$ or 4.
Fig.~\ref{probpart} shows the compatibility distribution
for the antihelium candidates (Fig.~\ref{probpart}a) and
helium samples together with
Monte Carlo predictions for the helium event distribution  
(Fig.~\ref{probpart}b).
As seen, the compatibility cut enables us to 
reject the small remaining background and
keep nearly all of the helium sample.

\begin{figure}[ht]
  \begin{center}
\begin{tabular}{c}                              % 
    \includegraphics[width=\figwidth ]{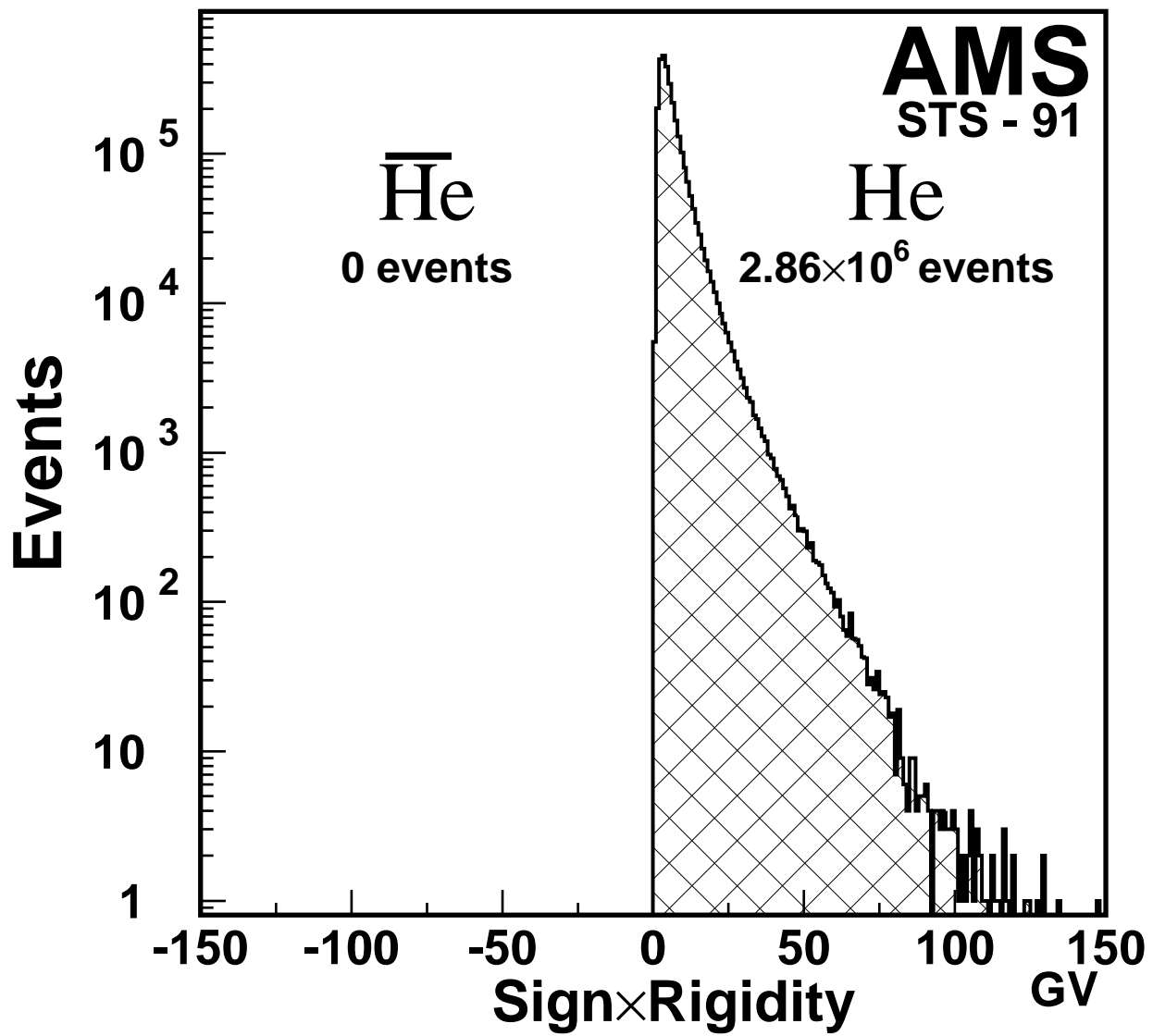} \\
\end{tabular}
  \end{center}
  \caption{Measured rigidity times the charge sign for 
          selected $|Z|=2$ events.}
  \label{finrige2}
\end{figure}

The results of our search are summarized in Fig.~\ref{finrige2}.
As seen, we obtain a total of 
$ 2.86 \times 10^6$ $\He$ 
events up to a rigidity of 140 GV.
We found no antihelium event at any rigidity.

\section*{Results and Interpretation}

Since no antihelium nuclei were observed,
we can only establish an upper limit on their flux.
Here three upper limits on this flux relative to the observed flux
of helium nuclei are calculated which differ 
in the assumptions used for the 
antihelium rigidity spectrum.
In the first it is assumed
to have the same shape as the helium rigidity spectrum.
In the second 
this spectrum is assumed to be uniform.
Finally a conservative estimate is made independent of the
antihelium rigidity spectrum.

All of these methods require the measured rigidity spectrum to be
corrected for the detector resolution 
and efficiency as a function of 
the measured, $\R_m$, and incident, $\R$, rigidity.
The detection efficiency
including the rigidity resolution function, 
$f(\R,\R_m)$,
was evaluated through complete Monte Carlo simulation using 
the GEANT Monte Carlo package~\cite{geant}.
The incident rigidity spectrum,  $dN'/d\R$
was extracted from the measured spectrum, $dN'/d\R_m$, 
by numerical deconvolution of 
$ dN'/d\R_m = \int (dN'/d\R) \times f(\R,\R_m)d\R$.
To obtain the detector efficiency for antihelium,
$\epsilon _\aHe(\R)$, 
a small correction was applied to
the efficiency for helium nuclei, $\epsilon _\He(\R)$,
based on the estimated~\cite{antihecorr}
difference in absorption cross sections.

Letting $N_\He(\R_i)$ be the number of incident
helium nuclei in the 
rigidity bin ($\R_i,\R_i+\Delta \R$) 
and $N'_\He(\R_i)$ be
the number of measured $\He$ in the same rigidity bin
after correction for the detector resolution,
then $N'_\He(\R_i) = \epsilon_\He(\R_i) N_\He(\R_i) $,
where $\epsilon _\He(\R_i)$ is the detection efficiency in this bin,
and similarly for antihelium.
Over the rigidity interval studied
no $\aHe$ were found, $N'_\aHe(\R_i) = 0$ for each $i$. 
At the 95\,\% confidence level this is taken to be less than 3  
and the differential upper limit for the flux ratio is given by:
\begin{equation}
\frac{N_\aHe(\R_i)}{N_\He(\R_i)} < \frac{3}{N'_\He(\R_i)} 
                                 \!\!\frac{\,\,/\epsilon_\aHe(\R_i)}
                                          {\,\,/\epsilon_\He(\R_i)}\:.
\label{eqdifflim}
\end{equation}
The difference between $\epsilon_\aHe(\R_i)$
and $\epsilon_\He(\R_i)$ is small,
so these terms practically cancel and
the results below are essentially independent of the detection efficiency.

\begin{figure}[ht]
  \begin{center}                    % 
    \includegraphics[width=\figwidth]{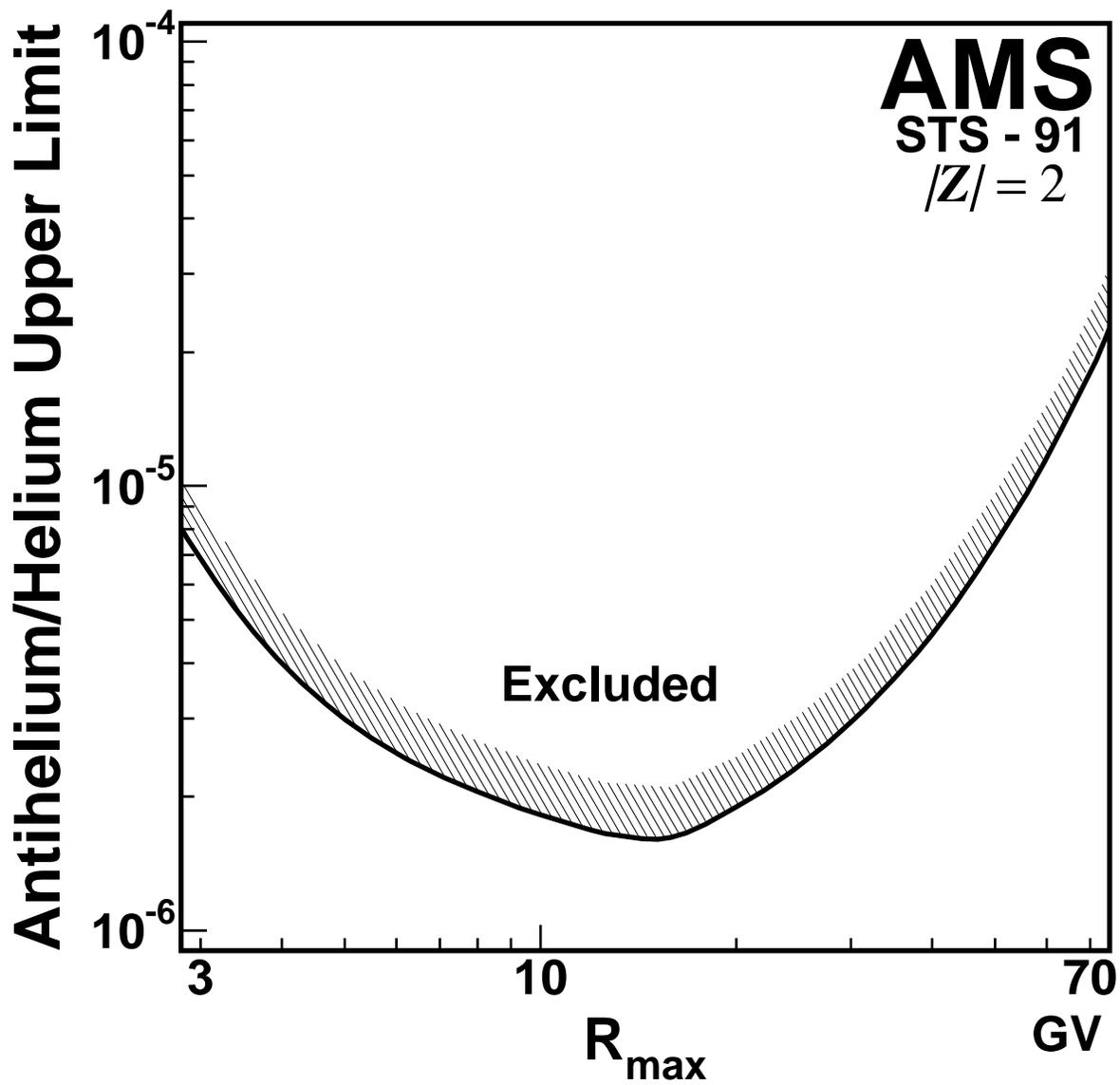} \\
  \end{center}
  \caption{Upper limits on the relative flux of antihelium to helium,
            at the 95\% confidence level,
            as a function of the rigidity interval
            $\R = 1.6\,\mathrm{GV}$ to $\R_{max}$. 
            These results are independent of the incident antihelium spectra.}
  \label{ahezeq2lim}
\end{figure}

\begin{itemize}
\item[(i)]
If the incident $\aHe$ spectrum is assumed to have the same shape as
the $\He$ spectrum over the range $1<\R<140$\,GV, then
summing equation~(\ref{eqdifflim}) yields a limit of:
\begin{eqnarray}
\nonumber\frac{N_\aHe}{N_\He} &<& 1.1 \times 10^{-6}.
\end{eqnarray}

\item[(ii)]
Assuming a uniform $\aHe$ rigidity spectrum, and using a
mean $\aHe$ inverse detection efficiency, 
$ \langle 1/\epsilon _\aHe\rangle= \sum (1/\epsilon _\aHe(\R_i))/n $,
and noting that $N'_\aHe = \sum N'_\aHe(\R_i) = 0$
which is also taken to be less than 3 at the 95\,\% C.L.,
summing equation~(\ref{eqdifflim}) yields a limit of
\begin{eqnarray}
\frac{N_\aHe}{N_\He}
= \frac{\sum N_\aHe(\R_i)}{\sum N_\He(\R_i)}
&<& \frac{3}{\sum N'_\He(\R_i)}
   \!\!\frac{\,\,\langle 1/\epsilon _\aHe\rangle}{\,\,/\epsilon_\He(\R_i)} ,
\\[2mm]
\nonumber
\hfill\mbox{which evaluates to\qquad }
\frac{N_\aHe}{N_\He}
  &<&
 1.8 \times 10^{-6} \mbox{ for } \R = 1.6 \mbox{ to } 40\,\mathrm{GV}\\
\nonumber\hfill\mbox{ and\qquad} 
\frac{N_\aHe}{N_\He}
  &<&3.9\times 10^{-6} \mbox{ for } \R = 1.6 \mbox{ to } 100\,\mathrm{GV}.
\label{intlim}
\end{eqnarray}
\item[(iii)]
For a conservative upper limit, which does not depend on the 
antihelium spectrum, 
equation~(\ref{eqdifflim}) is summed from $\R_{min}= 1.6$\,GV
up to a variable $\R_{max}$
and instead of the  mean value $\langle 1/\epsilon _\aHe\rangle$  
the minimum value of this efficiency in the $(\R_{min},\R_{max})$ 
interval is taken, yielding
\begin{equation}
\frac{\sum N_\aHe(\R_i)}{\sum N_\He(\R_i)} <
\frac{3}{\sum N'_\He(\R_i)}
\!\!\frac{\,\,/\epsilon _\aHe^{min}(\R_{min},\R_{max})}
         {\,\,/\epsilon _\He(\R_i)\hfill}
\mbox{, where } 
\R_i = (\R_{min},\R_{max}).
\label{conslim}
\end{equation}
These results are shown in Fig.~\ref{ahezeq2lim}\ as a
function of $\R_{max}$.
\end{itemize}

In conclusion, we found no antihelium nuclei at any rigidity. 
Up to rigidities of 140\,GV, 
$ 2.86 \times 10^6$ helium nuclei were measured.
Assuming the antihelium rigidity spectrum to
have the same shape as the helium spectrum,
an upper limit at the 95\,\% confidence level on the relative flux of
antihelium to helium of $1.1 \times 10^{-6}$ was obtained.
This result is an improvement in both sensitivity and rigidity range
over previous measurements~\cite{saeki98}.
This flight has shown that the completed AMS on the International
Space Station will provide many orders of magnitude of improvement
in the sensitivity to search for anithelium.

\newpage
\section*{Acknowledgements}
We thank Professors 
S.~Ahlen, 
C.~Canizares,
A.~De~Rujula,
J.~Ellis, 
A.~Guth,
M.~Jacob,
L.~Maiani,
R.~Mewaldt,
R.~Orava,
J.~F.~Ormes
and
M.~Salamon 
for helping us to initiate this experiment.

The success of the first AMS mission is due to
many individuals and organizations
outside of the collaboration.
The support of NASA
was vital in the inception, development and operation of the experiment.
The dedication of 
Douglas P.~Blanchard, 
Mark J.~Sistilli, 
James R.~Bates,
Kenneth Bollweg
and the NASA and Lockheed--Martin Mission Management team,
the support of the Max--Plank Institute for Extraterrestrial Physics,
the support of the space agencies from 
Germany (DLR),
Italy (ASI),
France (CNES) and
China 
and the support of CSIST, Taiwan, 
made it possible to complete this experiment on time.

The support of CERN and GSI--Darmstadt,
particularly of Professor Hans Specht and Dr.~Reinhard Simon
made it possible for us to calibrate the detector after the shuttle
returned from orbit.

We are most grateful to the STS--91 astronauts,
particularly to Dr.~Franklin Chang--Diaz 
who provided vital help to AMS during the flight.

The support of 
INFN, Italy, 
ETH--Z\"urich, 
the University of Geneva,
the Chinese Academy of Sciences,
Academia Sinica and National Central University, Taiwan,
the RWTH--Aachen, Germany,
the University of Turku, the University of Technology of Helsinki, Finland,
U.S.~DOE and M.I.T.
is gratefully acknowledged.

%
%%%%%%%%%%%%%%%%%%%%%%%%%%%%%%%%%%%%%%%%%%%%%%%%%%%%%%%%%%%%%%%%%%%%%%%%%%%%%%%
% Bibliography
%%%%%%%%%%%%%%%%%%%%%%%%%%%%%%%%%%%%%%%%%%%%%%%%%%%%%%%%%%%%%%%%%%%%%%%%%%%%%%
\newpage
%
% Style file to use with mcite.
% Use l3style with just cite.
%\bibliographystyle{../biblio/amsstylem}
%\bibliography{%
%antihelium.bib,%
%../biblio/amscollab,%
%../biblio/amsmisc,%
%../biblio/astro,%
%../biblio/misc%
%}
\bibliographystyle{amsstylem}
\bibliography{%
antihelium.bib,%
amscollab,%
amsmisc,%
astro,%
misc%
}

\begin{mcbibliography}{10}

\bibitem{smoot75}
G.F.Smoot \etal,
\newblock  \PRL {\bf 35}  (1975) 258--261\relax
\relax
\bibitem{steigman76}
G.Steigman \etal,
\newblock  Ann. Rev. Astr. Ap. {\bf 14}  (1976) 339\relax
\relax
\bibitem{badhwar78}
G.Badhwar \etal,
\newblock  Nature {\bf 274}  (1978) 137\relax
\relax
\bibitem{buffington81}
A.Buffington \etal,
\newblock  ApJ {\bf 248}  (1981) 1179--1193\relax
\relax
\bibitem{golden97}
R.L.Golden \etal,
\newblock  ApJ {\bf 479}  (1997) 992\relax
\relax
\bibitem{ormes97}
J.F.Ormes \etal,
\newblock  ApJ Letters {\bf 482}  (1997) L187\relax
\relax
\bibitem{saeki98}
T.Saeki \etal,
\newblock  \PL {\bf B422}  (1998) 319\relax
\relax
\bibitem{nimproposal}
S. Ahlen \etal,
\newblock  \NIM {\bf A350}  (1994) 351\relax
\relax
\bibitem{ams_00}
AMS Collab, J.~Alcaraz \etal,
\newblock  \NIM, ,
\newblock  in preperation\relax
\relax
\bibitem{viertel98}
\newline see also: G.~M.~Viertel, M.~Capell,
\newblock  \NIM {\bf A 419}  (1998) 295--299\relax
\relax
\bibitem{tracker}
AMS Tracker Group, G. Ambrosi \etal,
\newblock  \NIM (1999),
\newblock  to be submitted to NIM\relax
\relax
\bibitem{geant}
See R. Brun \etal, ``GEANT 3'', CERN DD/EE/84-1 (Revised), September 1987.\\
  The FLUKA program (see P.~A.~Aamio, FLUKA Users Guide, CERN Report
  TIS--RP--190 (1990)) is used to simulate hadronic interactions.\relax
\relax
\bibitem{antihecorr}
A.A.Moiseev and J.F.Ormes,
\newblock  Astroparticle Physics {\bf 6}  (1997) 379--386\relax
\relax
\end{mcbibliography}

%
%%%%%%%%%%%%%%%%%%%%%%%%%%%%%%%%%%%%%%%%%%%%%%%%%%%%%%%%%%%%%%%%%%%%%%%%%%%%%%
% Author List
%%%%%%%%%%%%%%%%%%%%%%%%%%%%%%%%%%%%%%%%%%%%%%%%%%%%%%%%%%%%%%%%%%%%%%%%%%%%%%
%
\newpage
\typeout{   }     
\typeout{Using author list for AMS paper 01 }
\typeout{$Modified: May 99 by M. Capell $}
\typeout{!!!!  This should only be used with document option a4p!!!!}
%
%
%
%  L A T E X  version!!
%
%
% Make sure that the Lep package has been used!
%\input{Lep.sty}%
%
%\ifx\LepCalled\undefined%
%\typeout{     }%
%\typeout{!!!!!!!!!!!!!!!!!!!!!!!!!!!!!!!!!!!!!!!!!!!!!!!!!!!!!!!!!!!}%
%\typeout{Yikes.  You haven't used the Lep package!}%
%\typeout{Please put \protect\usepackage\protect{Lep\protect} in your preamble,
%         followed by}%
%\typeout{\protect\Lep\protect{1\protect} or \protect\Lep\protect{2\protect}}%
%\typeout{     }%
%\typeout{For now you will get a Lep phase 2 authorlist (may not be right!).}%
%\typeout{!!!!!!!!!!!!!!!!!!!!!!!!!!!!!!!!!!!!!!!!!!!!!!!!!!!!!!!!!!!}%
%\typeout{     }%
%\Lep{2}\fi%

\newcounter{tutetotcount}
\newcounter{tutetutecount}
\newcounter{tutecount}
\newcounter{namecount}
\newcommand{\tutenum}[1]{%
\stepcounter{tutetotcount}%
\stepcounter{tutecount}%
\ifnum\value{tutecount}=27%
\stepcounter{tutetutecount}%
\addtocounter{tutecount}{-26}%
\fi%
\xdef#1{{\ifnum\value{tutetutecount}>0\alph{tutetutecount}\fi\alph{tutecount}}}}
\def\tute#1{$^{#1}$\stepcounter{namecount}}
\newcounter{notecount}
\newcommand{\note}{{\stepcounter{notecount}\thenotecount}}
%
%
% this will have to be reordered when the dust settles.
%
\tutenum\aachenI           % 1
\tutenum\aachenIII            % 1
\tutenum\lapp              % 4
\tutenum\jhu
\tutenum\lsu
\tutenum\cssa
\tutenum\calt
\tutenum\iee
\tutenum\ihep           % 7
\tutenum\bologna           % 9 
\tutenum\bucharest         % 12
\tutenum\mit               % 14 
\tutenum\ncu               % 48
\tutenum\coimbra
\tutenum\maryland
\tutenum\florence          % 15
\tutenum\mpi
%  \tutenum\cern              % 16 
\tutenum\geneva            % 18
\tutenum\grenoble
\tutenum\hut
\tutenum\ist
\tutenum\lip
\tutenum\csist
\tutenum\madrid            % 24
\tutenum\milan             % 25
\tutenum\kurch
\tutenum\moscow            % 26
\tutenum\perugia           % 31
\tutenum\as
% \tutenum\trento
\tutenum\korea
\tutenum\turku
\tutenum\eth               % 46
{
\parskip=0pt
\noindent
{\bf The AMS Collaboration:}
\ifx\selectfont\undefined%  old style font selection
 \baselineskip=10.8pt
 \baselineskip\baselinestretch\baselineskip
 \normalbaselineskip\baselineskip
 \ixpt
\else%                      new style font selection
% was {9}{10.8} but didn't fit. (206 auth, 31 inst, 9 notes, 7/99 mhc)
 \fontsize{9}{10.7pt}\selectfont
\fi
\medskip
\tolerance=10000
\hbadness=5000
\raggedright
\hsize=162truemm\hoffset=0mm
\def\r{\rlap,}
\noindent

J.Alcaraz\r\tute\madrid\
D.Alvisi\r\tute\bologna\
B.Alpat\r\tute\perugia\
G.Ambrosi\r\tute\geneva\
H.Anderhub\r\tute\eth\
L.Ao\r\tute\calt\
A.Arefiev\r\tute\moscow\
P.Azzarello\r\tute\geneva\
E.Babucci\r\tute\perugia\
L.Baldini\r\tute{\bologna,\mit}\
% ...
\typeout{--------------------------------------------------------------}
\typeout{
Imagine that:  <\thenamecount> authors from <\thetutetotcount> institutes.}
\typeout{--------------------------------------------------------------}
%
%\newpage
\vspace*{-\baselineskip}
\rule[\baselineskip]{\textwidth}{0.4pt}
\begin{list}{A}{\itemsep=0pt plus 0pt minus 0pt\parsep=0pt plus 0pt minus 0pt
                \topsep=0pt plus 0pt minus 0pt}
\item[$^\aachenI$]
 I. Physikalisches Institut, RWTH, D-52056 Aachen, Germany$^\note$
\item[$^\aachenIII$]
 III. Physikalisches Institut, RWTH, D-52056 Aachen, Germany$^6$
\item[$^\lapp$] Laboratoire d'Annecy-le-Vieux de Physique des Particules, 
     LAPP, F-74941 Annecy-le-Vieux CEDEX, France
\item[$^\lsu$] Louisiana State University, Baton Rouge, LA 70803, USA
\item[$^\jhu$] Johns Hopkins University, Baltimore, MD 21218, USA
\item[$^\cssa$] Center of Space Science and Application, 
  Chinese Academy of Sciences,
  100080 Beijing, China
\item[$^\calt$] Chinese Academy of Launching Vehicle Technology, CALT,
  100076 Beijing, China
\item[$^\iee$] Institute of Electrical Engineering, IEE, 
  Chinese Academy of Sciences, 100080 Beijing, China
\item[$^\ihep$] Institute of High Energy Physics, IHEP, 
  Chinese Academy of Sciences, 
  100039 Beijing, China$^\note$      
\item[$^\bologna$] University of Bologna and  INFN-Sezione di Bologna, 
     I-40126 Bologna, Italy 
\item[$^\bucharest$] Institute of Microtechnology, 
                    Politechnica University of Bucharest 
                    and University of Bucharest,
                    R-76900 Bucharest, Romania
\item[$^\mit$] Massachusetts Institute of Technology, Cambridge, MA 02139, USA
\item[$^\ncu$] National Central University, Chung-Li, Taiwan 32054
\item[$^\coimbra$]  Laboratorio de Instrumentacao e Fisica Experimental de 
            Particulas, LIP, P-3000 Coimbra, Portugal
\item[$^\maryland$] University of Maryland, College Park, MD 20742, USA
\item[$^\florence$] INFN Sezione di Firenze, 
     I-50125 Florence, Italy
\item[$^\mpi$] Max--Plank Institut fur Extraterrestrische Physik,
            D-85740 Garching, Germany
% \item[$^\cern$] European Laboratory for Particle Physics, CERN, 
%      CH-1211 Geneva 23, Switzerland
\item[$^\geneva$] University of Geneva, CH-1211 Geneva 4, Switzerland
\item[$^\grenoble$] Institut des Sciences Nucleaires,
     F-38026 Grenoble, France
\item[$^\hut$] Helsinki University of Technology,
    FIN-02540 Kylmala, Finland
\item[$^\ist$] Instituto Superior T\'ecnico, IST,  P-1096 Lisboa, Portugal
\item[$^\lip$] Laboratorio de Instrumentacao e Fisica Experimental de 
            Particulas, LIP, P-1000 Lisboa, Portugal
\item[$^\csist$] Chung--Shan Institute of Science and Technology,
     Lung-Tan, Tao Yuan 325, Taiwan 11529
\item[$^\madrid$] Centro de Investigaciones Energ{\'e}ticas, 
     Medioambientales y Tecnolog{\'\i}cas, CIEMAT, E-28040 Madrid,
     Spain$^\note$ 
\item[$^\milan$] INFN-Sezione di Milano, I-20133 Milan, Italy
\item[$^\kurch$] Kurchatov Institute, Moscow, 123182 Russia
\item[$^\moscow$] Institute of Theoretical and Experimental Physics, ITEP, 
     Moscow, 117259 Russia  
\item[$^\perugia$] INFN-Sezione di Perugia and Universit\'a Degli 
     Studi di Perugia, I-06100 Perugia, Italy$^\note$  
\item[$^\as$] Academia Sinica,
    Taipei, Taiwan
\item[$^\korea$] Kyungpook National University, 
     702-701 Taegu, Korea
%\item[$^\trento$] ITST Trento,
%    I-XXXXX Trento, Italy
\item[$^\turku$] University of Turku,
    FIN-20014 Turku, Finland
\item[$^\eth$] Eidgen\"ossische Technische Hochschule, ETH Z\"urich,
     CH-8093 Z\"urich, Switzerland
\setcounter{notecount}{0}
% notes in the authorlist
%
% J.Krieger, E. A. Werner 
\item[$^\note$] Now at ISATEC, Aachen, Germany.
%
% P. Levtchenko
\item[$^\note$] Permanent address: Nuclear Physics Institute, 
               St. Petersburg, Russia.
%
% G. Maehlum 
\item[$^\note$] Now at IDE--AS, Oslo, Norway.
%
% H. Postema
\item[$^\note$] Now at European Laboratory for Particle Physics, CERN, 
     CH-1211 Geneva 23, Switzerland.
%
% B. Verlaat
\item[$^\note$] Now at National Institute for High Energy Physics, NIKHEF, 
           NL-1009 DB Amsterdam, The Netherlands.
%
% noted in institute list
%
% for Aachen I & III
\item[$^\note$]  Supported by the 
Deutsches Zentrum f\"ur Luft-- und Raumfahrt, DLR.
%
% for IHEP
\item[$^\note$] Supported by the National Natural Science Foundation of China.
%
% for CIEMAT
\item[$^\note$] Supported also by the Comisi\'on Interministerial de Ciencia y 
           Tecnolog{\'\i}a.
%
% for Perugia
\item[$^\note$] Also supported by the Italian Space Agency.

\end{list}
}
\vfill

%%% Local Variables: 
%%% mode: latex
%%% TeX-master: t
%%% End:

%%% Local Variables: 
%%% mode: latex
%%% TeX-master: t
%%% TeX-master: t
%%% TeX-master: t
%%% End: 

%%% Local Variables: 
%%% mode: latex
%%% TeX-master: t
%%% End: 

%
%%%%%%%%%%%%%%%%%%%%%%%%%%%%%%%%%%%%%%%%%%%%%%%%%%%%%%%%%%%%%%%%%%%%%%%%%%%%%%
% Figures
%%%%%%%%%%%%%%%%%%%%%%%%%%%%%%%%%%%%%%%%%%%%%%%%%%%%%%%%%%%%%%%%%%%%%%%%%%%%%%
%
%
%The macro to make this figure (using Mn\_Fit\cite{mn_fit}) can be found in the
%directory {\tt /l3/paper/example/figs}. 

\newpage

\newpage

\newpage

\newpage

\newpage

\newpage

\newpage

\newpage

\newpage

\end{document}